\documentclass[sigconf]{acmart}

\copyrightyear{2026}
\acmYear{2026}

\acmConference[AgentSearch @ SIGIR '26]
{AgentSearch: The First Workshop on Indexing, Retrieval, and Ranking of AI Agents, co-located with the 49th International ACM SIGIR Conference on Research and Development in Information Retrieval}
{July 24, 2026}
{Melbourne | Naarm, Australia}

\author{Miaomiao Cai}
\affiliation{%
  \institution{National University of Singapore}
  \city{Singapore}
  \country{Singapore}
  }
\email{cmm.hfut@gmail.com}

\author{He Chang}
\affiliation{%
  \institution{Communication University of China}
  \city{Beijing}
  \country{China}
  }
\email{hechangcuc@cuc.edu.cn}

\author{Yunshan Ma}
\affiliation{%
  \institution{Singapore Management University}
  \city{Singapore}
  \country{Singapore}
  }
\email{ysma@smu.edu.sg}

\author{See-Kiong Ng}
\affiliation{%
  \institution{National University of Singapore}
  \city{Singapore}
  \country{Singapore}
  }
\email{seekiong@nus.edu.sg}

\begin{document}

\settopmatter{authorsperrow=4}

\renewcommand{\shortauthors}{Miaomiao Cai et al.}

\title{ForecastAgentSearch: Towards a Multi-Expert Agent Search System for Geopolitical Event Forecasting}

\maketitle

\section{Extended Abstract}

Geopolitical event forecasting aims to anticipate future political, military, and social developments from historical observations and evolving real-world contexts~\cite{zhao2021event,deng2024advances,ma2023context}. 
It is important for early warning, risk assessment, policy planning, and strategic decision-making in high-impact scenarios such as international conflicts, diplomatic actions, sanctions, protests, humanitarian crises, and regional instability~\cite{zhao2021event,jin2021forecastqa,halawi2024approaching}. 
This task is especially challenging in regions such as the Middle East, where future events are often shaped by intertwined factors, including regional power dynamics, political alliances, economic pressure, religious and cultural tensions, historical grievances, and multimodal media narratives~\cite{thinktankme,mmforecast}. 
Therefore, effective forecasting requires not only temporal reasoning over historical events, but also the ability to identify which evidence and expertise are useful for a specific forecasting task.

Recent advances in large language models (LLMs) have created new opportunities for event forecasting. 
LLM-based methods can process textual contexts, retrieve relevant historical evidence, and generate predictions through prompting, in-context learning, chain-of-thought reasoning, or retrieval-augmented generation~\cite{liao2024gentkg,luo2024chain,chang2024comprehensive,lewis2020retrieval}. 
However, most existing approaches still rely on either a single general-purpose predictor or a fixed expert ensemble. 
A single predictor may follow a dominant reasoning path and overlook alternative geopolitical perspectives, while a fixed expert ensemble may introduce redundant, irrelevant, or costly expert opinions~\cite{ye2024mirai,thinktankme,cai2024survey}. 
Such designs are insufficient for complex geopolitical forecasting, where different queries may require different combinations of regional, political, economic, religious, historical, multimodal, and risk-oriented expertise.

Recent forecasting studies suggest two important observations. 
First, heterogeneous evidence sources may play different functional roles in prediction. 
For example, multimodal evidence can either highlight salient historical events or provide complementary context beyond textual descriptions~\cite{mmforecast}. 
Second, specialized expert models can provide complementary predictive knowledge, but their usefulness is often query-dependent~\cite{thinktankme,cai2024survey}. 
These observations indicate that reliable geopolitical forecasting requires more than stronger predictive models or larger context windows. 
It also requires a principled mechanism for deciding \emph{which experts should be consulted, how they should be ranked, and how their outputs should be coordinated}.

To this end, we introduce \textbf{ForecastAgentSearch}, a system-level formulation that treats complex geopolitical event forecasting as a \emph{multi-expert agent search problem}. 
Rather than directly mapping historical contexts to predictions, ForecastAgentSearch introduces an intermediate search and coordination layer over specialized expert agents. 
In this formulation, expert agents are treated as searchable, rankable, and composable resources, which naturally connects geopolitical forecasting with core problems in information retrieval and agent search~\cite{wu2026agentsearchbench,shi2025retrieval,braunschweiler2025toolreagt}.

\begin{figure*}[t]
    \centering
    \includegraphics[width=\textwidth]{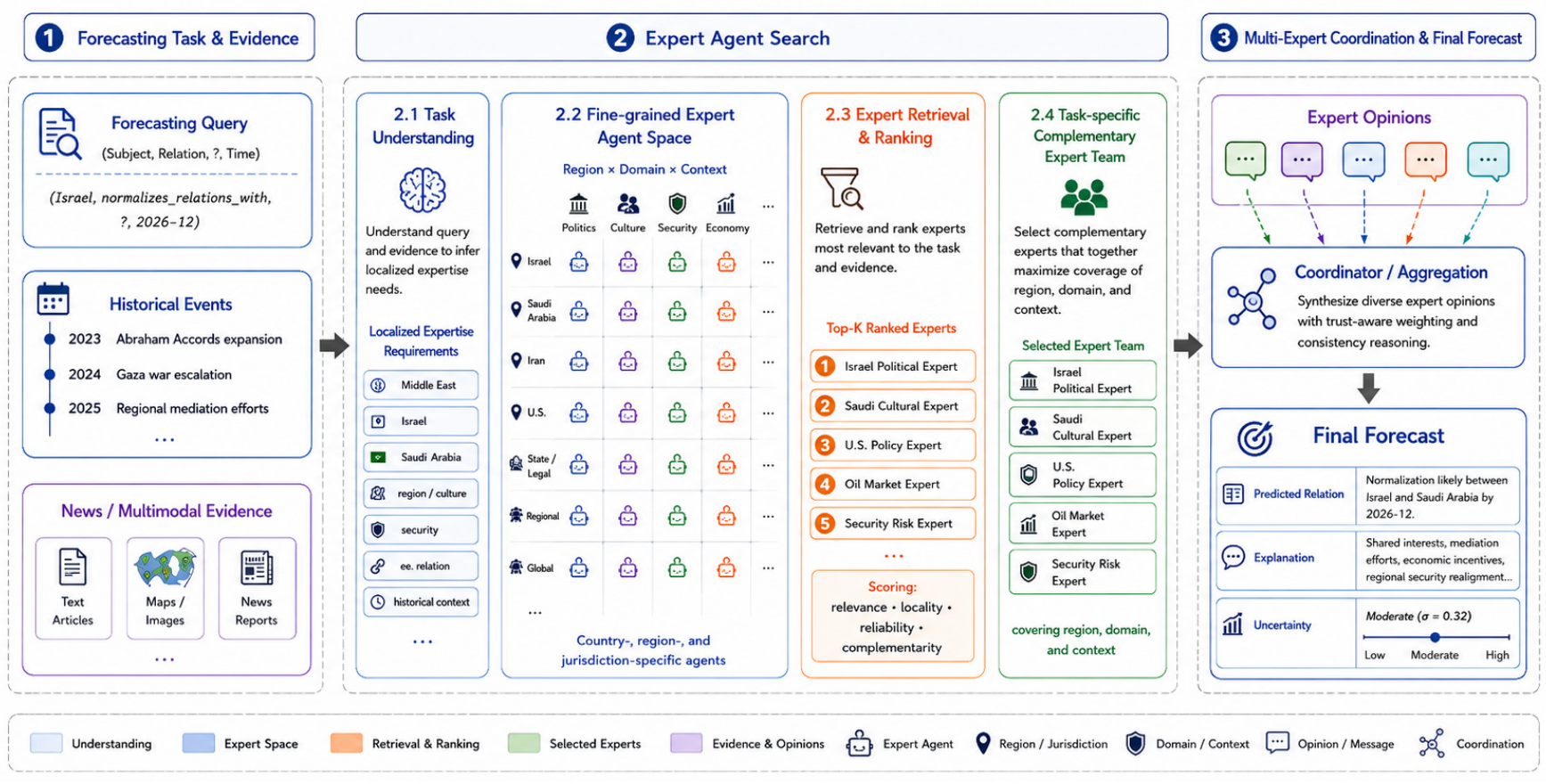}
    \caption{Overview of \textbf{ForecastAgentSearch}. Given a geopolitical forecasting query and heterogeneous evidence, the system first identifies task-specific expertise requirements, then searches over a fine-grained expert agent space, retrieves and ranks the most suitable agents, and coordinates their outputs to generate the final forecast with explanations and uncertainty estimates. }
    \label{fig:overview}

\end{figure*}

As illustrated in Figure~\ref{fig:overview}, ForecastAgentSearch consists of three main stages. 
The first stage takes a forecasting task and heterogeneous evidence as input, including the forecasting query, historical events, news context, and multimodal evidence. 
These sources provide complementary signals for prediction: historical events capture temporal actor interactions, news reports provide contextual narratives, and multimodal evidence may reveal additional regional or situational information. 
Instead of treating all evidence uniformly, ForecastAgentSearch performs task understanding to infer the expertise requirements behind the current query, such as regional locality, political relations, economic pressure, religious or cultural factors, security risks, and historical context.

The second stage searches over a fine-grained expert agent space.
Each expert agent is associated with a lightweight profile that describes its region or actor specialization, domain expertise, supported evidence sources, reliability estimate, inference cost, and known limitations.
For example, a forecasting query may require experts on Israeli political dynamics, Saudi cultural factors, U.S. policy, oil markets, religious tensions, multimodal conflict evidence, or regional security risks.
These profiles allow expert agents to be indexed and retrieved according to both structured metadata and semantic descriptions.

Given a forecasting query, ForecastAgentSearch retrieves and ranks candidate agents according to several task-aware criteria, including relevance to the query, regional or actor locality, historical reliability, inference cost, and complementarity with other selected experts.
Rather than consulting all available agents, the system selects a compact set of top-ranked experts that can provide complementary perspectives for the current task.
This design reflects the nature of geopolitical forecasting: the usefulness of an expert is highly dependent on the specific event, region, actors, and evidence sources involved.

The final stage coordinates the selected experts to produce the forecast.
Each expert can contribute a prediction, an intermediate analysis, supporting evidence, a confidence estimate, or possible risk factors from its own perspective.
A coordination module then synthesizes these outputs into the final forecast, together with explanations and uncertainty signals.
When experts disagree, the coordinator can compare their evidence, reliability, and domain coverage, rather than simply averaging their predictions.
This process resembles a structured think-tank workflow, where multiple specialists contribute different views and a coordinator synthesizes them into a coherent judgment.

The main contribution of this extended abstract is threefold. 
First, we formulate complex geopolitical event forecasting as an expert-agent search problem, shifting the focus from prediction alone to task-aware expert selection and coordination. 
Second, we outline \textbf{ForecastAgentSearch}, a system framework that retrieves, ranks, and coordinates specialized agents with different forms of expertise, including regional, political, economic, religious, historical, multimodal, and risk-oriented knowledge. 
Third, we discuss Middle East event forecasting as a representative testbed for this formulation, due to its complex regional interactions, heterogeneous evidence sources, and diverse analytical perspectives.

Although this extended abstract focuses on the problem formulation and system design, ForecastAgentSearch naturally suggests several evaluation directions. Forecasting quality can be measured by accuracy, Brier score, log score, calibration error, temporal generalization, and uncertainty quality. The quality of agent search can be evaluated by determining whether retrieved experts match the required regions, domains, actors, and evidence sources. Future ablations can compare task-aware agent search with single-predictor forecasting, fixed expert ensembles, all-expert consultation, random expert selection, and variants without reliability-, cost-, or complementarity-aware ranking.

\section{Conclusion}

This extended abstract introduces \textbf{ForecastAgentSearch}, which formulates complex geopolitical event forecasting as a multi-expert agent search problem. 
Instead of relying on a single predictor or a fixed expert ensemble, ForecastAgentSearch treats expert agents as searchable and composable resources, and aims to retrieve, rank, and coordinate suitable experts according to task-specific requirements.

Motivated by recent findings that both evidence utility and expert usefulness are task-dependent, ForecastAgentSearch highlights the need for structured expert profiling, task-aware retrieval, reliability- and cost-aware ranking, complementary team formation, and interpretable aggregation. 
Middle East event forecasting serves as a meaningful testbed, since it requires reasoning over regional, political, economic, religious, historical, and multimodal factors.

\bibliographystyle{ACM-Reference-Format}
\bibliography{sample-base}

\end{document}